\documentclass[twocolumn,showpacs,preprintnumbers,amsmath]{revtex4}
\usepackage{amssymb}

\usepackage{graphicx}

\tighten 

\begin{document}

\title{Topological Gauge Structure and Phase Diagram of a Doped Antiferromagnet}
\author{Su-Peng Kou$^1$ and Zheng-Yu Weng$^2$}
\address{$^1$Department of Physics, Beijing Normal University, Beijing, 100875, China\\
$^2$Center for Advanced Study, Tsinghua University, Beijing, 100084, China}

\begin{abstract}
We show that a topological gauge structure in an effective description of
the \textrm{t-J} model gives rise to a global phase diagram of
antiferromagnetic (AF) and superconducting (SC) phases in a weakly doped
regime. Dual confinement and deconfinement of holons and spinons play
essential roles here, with a quantum critical point at a doping
concentration $x_c\simeq 0.043$. The complex experimental phase diagram at
low doping is well described within such a framework.
\end{abstract}

\maketitle

\draft



\emph{Introduction. }Cuprate superconductors have shown different ordering
tendencies as the hole concentration $x$ varies \cite{keimer,chou,CN,ino}.
The parent state is a Mott insulator with an AF long range order (AFLRO).
Hole doping leads to the disappearance of AFLRO at $x_0\sim 0.02.$ In a
lightly doped region, $x_0<x<x_c\sim 0.05,$ the low temperature (T) state is
a cluster spin-glass phase with localized holes. At $x>x_c,$ the holes are
delocalized and the ground state becomes a d-wave SC phase.

In literature, based on the three-band model, dipole defects induced by
holes have been conjectured \cite{aharony,glazman,che} as responsible for
the destruction of the AFLRO as well as the spin-glass phase \cite{good1}. A
different cause for destroying the AFLRO has been attributed \cite{timm} to
holes dressed with vortices \cite{meron}. But a systematic model study is
still lacking, which prevents us to fully understand the evolution of the AF
and SC phases at low doping.

Theoretically, the single-band $t-J$ model has been widely used to describe
electronic properties in cuprates. But it is notably difficult to conduct a
reliable investigation continuously going from half-filling to a
superconducting phase. Thus, a tractable model of the doped AF-Mott
insulator is called for. Such a model should well describes
antiferromagnetism at half-filling, on one hand, and give rise to a d-wave
superconductivity at large doping, on the other hand, while the Hilbert
space remains restricted with $x$ as the \emph{natural} charge carrier
concentration.

An effective description which satisfies the above criteria has been derived 
\cite{string2} from the $\mathrm{t-J}$ Hamiltonian based on the bosonic
resonating-valence-bond (b-RVB) pairing \cite{lda,aa}. It is given by $%
H_{string}=H_h+H_s$ with 
\begin{eqnarray}
H_h &=&-t_h\sum_{\langle ij\rangle }(e^{iA_{ij}^s-i\phi
_{ij}^0})h_i^{\dagger }h_j+H.c.  \label{hh} \\
H_s &=&-J_s\sum_{\langle ij\rangle \sigma }(e^{i\sigma A_{ij}^h})b_{i\sigma
}^{\dagger }b_{j-\sigma }^{\dagger }+H.c.  \label{hs}
\end{eqnarray}
Here $h_i$ and $b_{i\sigma }$ are \emph{bosonic} holon and spinon operators,
respectively. At half-filling, $H_h$ is absent while $H_s$ reduces to the
Schwinger-boson mean-field Hamiltonian \cite{aa} ($A_{ij}^h=0),$ which
correctly characterizes AF correlations. In particular, a \emph{spinon} Bose
condensation, $<b_{i\sigma }>\neq 0,$ leads to an AFLRO. On the other hand,
when \emph{holons} are Bose condensed at \emph{large} doping, the ground
state becomes a d-wave SC \cite{string2,zhou}. \emph{Without} considering
the spinon condensation, such a phase would extrapolate at $x\rightarrow 0$
with $T_c\rightarrow 0$. But at low doping, spins can still remain ordered$.$
It thus may push the SC phase boundary to a finite $x_c,$ leaving a region
for a rich competing phenomenon.

In this Letter, we explore such a regime lying between the half-filling
antiferromagnetism and the superconducting phase. We find that the
theoretical phase diagram bears striking similarities to the experimental
one, and is controlled by a dual confinement-deconfinement procedure
determined by the topological gauge structure in (\ref{hh}) and (\ref{hs}).

\emph{Topological gauge structure.} The nontriviality of $H_{string}$ arises
from the link variables, $A_{ij}^s$ and $A_{ij}^h,$ in the doped case. These
link variables satisfy topological conditions: $\sum_cA_{ij}^s=\pm \pi
\sum_{l\in c}(n_{l\uparrow }^s-n_{l\downarrow }^s)$ and $\sum_cA_{ij}^h=\pm
\pi \sum_{l\in c}n_l^h$ for a closed loop $c$ (here $n_{l\sigma }^s$ and $%
n_l^h$ denote spinon and holon number operators, respectively). So $A_{ij}^s$
reflects the frustrations of spin background on the kinetic energy of charge
degrees of freedom, while $A_{ij}^h$ represents the influence of charge part
on spin degrees of freedom. [$\phi _{ij}^0$ in (\ref{hh}) describes a
uniform $\pi $ flux per plaquette with $\sum_{\square }\phi _{ij}^0=\pm \pi
] $.

It has been previously shown \cite{zhou} that in the superconducting state
with a holon Bose condensation, spinons are ``confined'', due to the gauge
field $A_{ij}^s$, to form integer spin excitations and nodal quasiparticles
emerge as recombined holon-spinon composites. The deconfinement of spinon
pairs occurs at $T\geq T_c$, which is responsible for destroying the SC\
phase coherence$.$ In the following, we show that when spinons are
condensed, holons will be ``confined'' too, due to the gauge field $A_{ij}^h$%
.

\emph{Holon confinement.} If spinons are condensed, it is straightforward to
observe that a holon will cost a logarithmically divergent energy in $H_s$.
Using an expansion $e^{i\sigma A_{ij}^h}\simeq 1+i\sigma
A_{ij}^h-(A_{ij}^h)^2/2,$ the energy cost of a holon in $H_s$ can be
estimated as 
\begin{equation}
\Delta E_s\sim J_s<b^{\dagger }><b^{\dagger }>\sum_{\left\langle
ij\right\rangle }(A_{ij}^h)^2\sim J_s\mathrm{ln}({L/a)},  \label{me}
\end{equation}
where $L$ is the size of the sample and $a$ is lattice constant. Such an
infinite energy can be physically understood as follows. In terms of the
spin flip operator \cite{string2} 
\begin{equation}
S_i^{+}=(-1)^ib_{i\uparrow }^{\dagger }b_{i\downarrow }\exp \left[ i\Phi
_i^h\right] ,
\end{equation}
the spin polarization $\left\langle S_i^{+}\right\rangle $ is twisted away
from the N\'{e}el order $(-1)^i<b_{i\uparrow }^{\dagger }><b_{i\downarrow }>$
by an angle $\Phi _i^h$ $=\sum_{l\neq i}\mbox{Im}\ln $ $(z_i-z_l)$ $n_l^h$,
which has a ``meron'' configuration as $\Phi _i^h\rightarrow \Phi _i^h\pm
2\pi $ by going around a holon once. In other words, a holon defined here is
a topological object carrying a spin-meron twist$.$ \emph{So a single holon
will not appear in the low-energy spectrum and must be confined in the spin
ordered phase.} 
\begin{figure}[tbp]
\begin{center}
\includegraphics{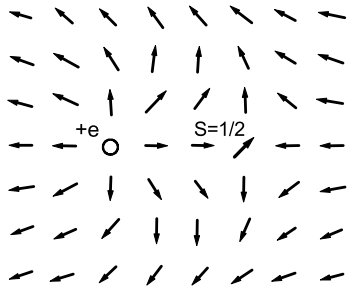}
\end{center}
\caption{In the spinon Bose condensed phase, a holon-meron is confined with
an antimeron to form a hole-dipole composite, which carries a charge +e and
spin 1/2. The arrows denote $\mathbf{n}_i$ $\varpropto $ $\mathbf{(-}1%
\mathbf{)}^i\left\langle \mathbf{S}_i\right\rangle $}
\label{fig1}
\end{figure}

It is easy to show that two holons will \emph{repulse} each other
logarithmically in $H_s$ as they are merons of the \emph{same} topological
charge \cite{kou}. The only way to realize a finite-energy \emph{hole}
object in the present framework is for each holon-meron to ``nucleate'' an
anti-meron from the vacuum. Such an anti-meron corresponds to a twist 
\begin{equation}
<b_{i\sigma }>\rightarrow <\bar{b}_{i\sigma }>\exp \left[ i\frac \sigma 2%
\mbox{Im}\ln (z_i-z_0)\right] ,
\end{equation}
with $z_0$ denoting its position. Then, the meron twist $\Phi _i^h$ in $%
\left\langle S_i^{+}\right\rangle $ or the link variable $A_{ij}^h$ in $%
\Delta E_s$ will be canceled out by the corresponding anti-merons if $z_0$'s
approaches to the holon positions. Each doped hole will then behave like a 
\emph{dipole} composed of a confined pair of meron (holon) and anti-meron as
shown in Fig. 1.

The total energy of a dipole can be generally written by \cite{kou}

\begin{equation}
E_{dipole}=E_{\text{core}}^h+E_{\text{core}}^m+V,  \label{edipole}
\end{equation}
where the infinite energy in Eq. (\ref{me}) is replaced by a finite one 
\begin{equation}
V(|\delta \mathbf{R|})=\Delta E_s\sim J_s\ln (|\delta \mathbf{R|}/a),
\label{v}
\end{equation}
in which $\delta \mathbf{R}$ denotes the spatial separation of two poles of
the dipole. Here $E_{\text{core}}^h$ and $E_{\text{core}}^m$ represent,
respectively, the core energies of a holon-meron and an anti-meron [$E_{%
\text{core}}^h$ also includes the hopping energy from (\ref{hh})]$.$

Therefore, a holon must be confined to an anti-meron to form a hole-dipole
object in the spin ordered phase. Note that a dipolar result was first found
by Shraiman and Siggia \cite{ss} based on a semi-classical treatment of the $%
t-J$ model. But there is an important difference: in the present case the
hole sits at a \emph{pole} instead of the \emph{center} of a dipole as shown
in Fig. 1. A more detailed and elaborate discussion on physical properties
of hole-dipoles in the spinon condensed phase is given elsewhere \cite{kou}.

\emph{Quantum critical point (QCP) }$x_{c\text{. }}$ The holon confinement
holds only at low doping. With the increase of $x$, as more and more
hole-dipoles are present at $T=0$, the ``confining'' potential (\ref{v}) can
get fully screened for large pairs of meron and anti-meron, leading to a
topological transition at $x=x_c,$ from dipoles to free merons in a fashion
of the Kosterlitz-Thouless (KT) transition \cite{BKT}.

We shall employ a standard KT\ renormalization group (RG) method to
calculate $x_c$. For this purpose we rewrite interaction $V(r)$ in (\ref
{edipole}) as $2\pi \beta ^{-1}K\ln (r/a),$ where the reduced spin stiffness 
$K(a)\varpropto $ $J_s\beta $ ($\beta =1/T$)$.$ The probability of creating
a meron-antimeron pair with two poles separated by a distance $a$ is given
by the pair fugacity $y^2(a)$. Note that in the conventional KT theory, $%
y^2(a)=e^{-\beta (E_{\text{core}}^h+E_{\text{core}}^m)}$, but in the present
case the dipole number is \emph{fixed} at $x$ per site and thus the initial $%
y^2(a)$ must be adjusted accordingly (see below).

In the RG scheme, small pairs of sizes within $r$ and $r+dr$ are integrated
out starting from\textbf{\ }the lattice constant $r=a$. The renormalization
effect is then represented by renormalized quantities $X(r)\equiv \frac
1{K\left( r\right) }$ and $y^2(r)$, which satisfy the famous recursion
relations \cite{BKT,cha} 
\begin{eqnarray}
dy^2/dl &=&2(2-\frac \pi X)\,y^2,  \label{rc1} \\
dX/dl &=&4\pi ^3y^2,  \label{rc2}
\end{eqnarray}
where $r=ae^l.$ What makes the present approach different from the
conventional KT theory is the presence of a finite density of the
hole-dipoles even at $T=0$ as pointed out above$.$ Here, by noting $\frac{%
y^2(r)}{r^4}d^2\mathbf{r}$ as the areal density of pairs of sizes between $r$
and $r+dr$ \cite{BKT}, we have the following constraint 
\begin{eqnarray}
x/a^2 &=&\int_a^\infty dr\,2\pi r\,\frac{y^2(r)}{r^4}  \nonumber \\
&=&\frac 1{2\pi ^2a^2}\int_0^\infty dle^{-2l}\,\frac{dX}{dl}\text{ ,}
\label{rc3}
\end{eqnarray}
In obtaining the second line the recursion relations (\ref{rc1}) and (\ref
{rc2}) are used.

The RG flow diagram of (\ref{rc1}) and (\ref{rc2}) is well known: the two
basins of attraction are separated by the initial values which flow to $%
X^{*}=\pi /2$ and $y^{*}=0$ in the limit $l\to \infty $. For $T\to 0,$ $%
X(l=0)\rightarrow 0$, the separatrix of the RG flows is given by 
\begin{equation}
l=\int_0^X\frac{dX^{\prime }}{4(X^{\prime }-\pi /2)-2\pi \ln (2X^{\prime
}/\pi )},  \label{rc4}
\end{equation}
and the critical hole density can be numerically determined in terms of (\ref
{rc3}) and (\ref{rc4}) as 
\begin{equation}
x_c\simeq \frac{0.84}{2\pi ^2}=0.043.  \label{xc}
\end{equation}
Therefore, a QCP is found at $x_c$ where hole-dipoles dissolve into
holon-merons and antimerons. At $x<x_c$, since the fugacity $y$ is always
renormalized to zero, each holon-meron has to be bound to an immobile
antimeron, implying that the doped holes should be self-trapped in space 
\cite{kou}.

\emph{Disappearance of AFLRO.} Even in the confined phase, $x<x_c$, the
system is not necessarily always AF ordered. The presence of hole-dipole can
lead to the destruction of the true AFLRO before reaching $x_c$. The basic
physics reason is due to the fact that dipoles have a long-range effect ($%
\frac 1r$) on the distortion of the magnetization direction.

Let us introduce a unit N\'{e}el vector $\mathbf{n}_i$ $\varpropto $ $%
\mathbf{(-}1\mathbf{)}^i\left\langle \mathbf{S}_i\right\rangle .$ Define $%
n_i^x+in_i^y\equiv e^{i\phi _i+i\phi _0}$, with $\mathbf{n}_0=$ $(\cos \phi
_0,\sin \phi _0)$ as the global AF magnetization direction$.$ Then a
hole-dipole centered at the origin [Fig. 1] will give rise to 
\begin{equation}
\phi _i\simeq (\mathbf{p\cdot r}_i)/|\mathbf{r}_i|^2  \label{dipole}
\end{equation}
at $|\mathbf{r}_i|>>|\mathbf{p|,}$ with $\mathbf{p\equiv -\hat{z}\times }%
\delta \mathbf{R}$ .

The twist of the N\'{e}er vector due to a dipole at the original is given by 
$\delta \mathbf{n}_i\mathbf{=n}_i\mathbf{-n}_0\approx \mathbf{m}\phi _i,$
with the unit vector $\mathbf{m}=(-\sin \phi _0$, $\cos \phi _0).$ In the
continuum limit, $\nabla ^2\mathbf{n(r)}=2\pi \mathbf{m}\left( \mathbf{%
p\cdot \nabla }\right) \delta (\mathbf{r}).$ The multi-dipole solution can
be generally written as $\nabla n^\mu \mathbf{(r)}=2\pi \sum_l\mathbf{p}%
_l\delta (\mathbf{r-r}_l)m_{l\text{ }}^\mu +\mathbf{g}_{\bot }^\mu $ where $%
l $ denotes the index of hole-dipoles and $\nabla \cdot \mathbf{g}_{\bot
}^\mu =0$. The transverse component $\mathbf{g}_{\bot }^\mu $ will have no
effect in the nonlinear sigma model \cite{che} and we may only focus on the
longitudinal part of $\nabla n^\mu \mathbf{(r)}$ below. At low temperature,
we may assume that the hole-dipoles are localized and treat all the
variables, $\mathbf{m}_l,$ $\mathbf{p}_l$, and $\mathbf{r}_l,$ as quenched.
Defining the quenched average $\left\langle \cdot \cdot \cdot \right\rangle
_q,$\emph{\ }and using $\left\langle m_l^\mu m_{l^{^{\prime }}}^\nu
\right\rangle =1/2\delta _{\mu \nu }\delta _{ll^{^{\prime }}},$ $%
\left\langle p_l^ip_{l^{^{\prime }}}^j\right\rangle =\delta _{ij}\delta
_{ll^{^{\prime }}}\eta a^2/2$ with $\eta =\left\langle |\delta \mathbf{R}%
|^2\right\rangle _q/a^2,$ we get 
\begin{equation}
\left\langle \partial _in^\mu (\mathbf{r})\partial _j^{^{\prime }}n^\nu (%
\mathbf{r}^{\prime })\right\rangle _q=\upsilon \delta _{ij}\delta _{\mu \nu
}\delta (\mathbf{r}-\mathbf{r}^{\prime })  \label{quench}
\end{equation}
in which $\upsilon =Ax$, with $A=\pi ^2\eta $.

The RG study of the non-linear sigma model with quenched random dipole
moments has been given \cite{che} within a one-loop approximation. Even
though the origin of the dipole moments is different, once (\ref{quench}) is
determined, these results can be directly applied here. The AF correlation
length $\xi $ has been obtained at low T as $\xi /a\simeq \exp (\frac{2\pi }{%
3\upsilon })$. The N\'{e}el temperature $T_N(x)$ is roughly given by the
solution of $\alpha \xi ^2\simeq a^2,$where $\alpha \sim 10^{-5}$ \cite
{keimer}, representing the effect of the interlayer coupling $J_{\perp }/J$.
Then the critical doping $x_0$ at which the AFLRO disappears can be
estimated by $x_0=-\frac{4\pi }{3A\ln \alpha }$. In order to get the
experimental value $x_0\sim 0.02,$ it has been assumed $A\sim 20$ in Ref. 
\cite{che}. In the present case, a self-consistent calculation leads to $%
\eta =1.23$ and $x_0\simeq 0.03$, determined by the KT theory based on (\ref
{rc1}) and (\ref{rc2}). On the other hand, at $x\rightarrow 0$ where $%
\upsilon \ll t\equiv T_N/\rho _s$ ($\rho _s\sim 0.176J$ is the spin
stiffness \cite{aa}), $\xi /a\sim \exp (\frac{2\pi }{3\upsilon }[1-(1-\frac
\upsilon t)^3])$ \cite{che} and one obtains $T_N(x)\approx T_N(0)-Ax\rho _s$%
, with $T_N(0)=-\frac{4\pi }{\ln \alpha }\rho _s$. The plot of $T_{N\text{ }%
} $ as a function of $x$ is shown in Fig. 2.

In Fig. 2, a characteristic temperature $T_f$ in the AFLRO phase is also
shown by the dotted curve, which represents the fact that although the holes
are all localized, the directions of their dipole moments can still rotate
freely to reach annealed equilibrium above $T_f.$ The dipole--dipole
interaction causes an energy difference of two dipoles, from parallel to
perpendicular in their relative moment alignment, is proportional to $1/r^2$
($r$ is the spatial separation between them). Associating $r$ with the
average hole-hole distance: $r=a/\sqrt{x}$, the interaction energy then
scales linearly with $x$, such that $T_f\sim 1/r^2\sim x$ \cite{aharony,good}%
. In the region $x_0<x<x_c,$ the AFLRO is destroyed and the AF orders are
limited mainly by finite size effects, where the size of the AF domains is
determined by hole concentration, $\xi \sim a/\sqrt{x}.$ The spin-glass
freezing temperature is then expected to vary as $T_g\sim \xi ^2\sim 1/x.$
Such a phase has been known as a cluster spin glass \cite{good1}. Below the
temperature $T_g$ the hole-dipolar configurations form a glass and their
dynamics strongly slows down.

\emph{Superconducting phase}. At $x\geq x_c,$ holons are deconfined and
free, and thus will experience Bose condensation at low T, giving rise \cite
{string2,zhou} to a d-wave superconducting ground state. 
\begin{figure}[tbp]
\includegraphics{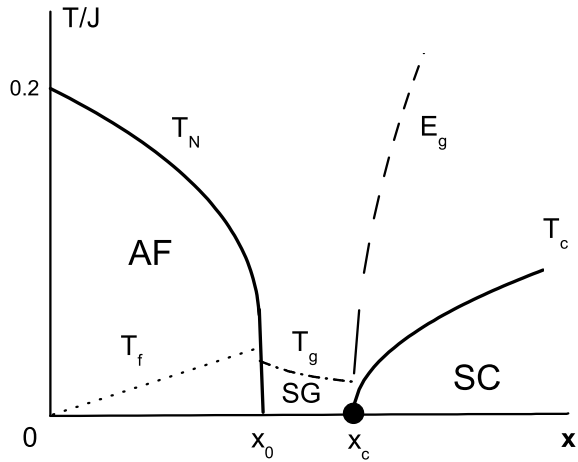}
\caption{Phase diagram at low doping $x$ : a dual confinement-deconfinement
occurs at a quantum critical point $x_c\simeq 0.043$. The N\'{e}el
temperature $T_N$ vanishes at $x_0\simeq 0.03$. $T_f$ and $T_g$ denote
characteristic spin freezing temperatures, and $T_c$ is the superconducting
transition temperature. $E_g$ is a characteristic spin energy (see text).}
\label{fig2}
\end{figure}

We have shown that the holons are confined to form immobile dipoles at $%
x\leq x_c$, instead of being Bose condensed. At $x=x_c,$ the concentration
of anti-merons is $x_c$, while at $x\gtrsim x_c,$, for each additional
holon, there will be no more anti-meron to be created so that the
concentration of anti-merons is roughly fixed at $x_c$ until $<\mid
b_{i\sigma }\mid >=0$. Thus, in an overlap regime of $<\mid b_{i\sigma }\mid
>\neq 0$ and $<h_i>\neq 0$ at $x\gtrsim x_c$, $A_{ij}^h$ in $H_s$ should be
replaced by a $\tilde{A}_{ij}^h$ due to the presence of antimerons, and its
strength $\tilde{B}^h=\pi x_{eff}/a^2$ is now controlled by an effective
concentration $x_{eff}=x-x_c$, instead of $x$ itself. Eventually at higher
doping, $<\mid b_{i\sigma }\mid >=0$ with no more (anti)merons, one will
recover $x$ from $x_{eff}$ as the parameter representing the doping effect
on the SC properties \cite{string2}.

At small doping, the continuum version of (\ref{hs}) is a $\mathrm{CP}^1$ 
\cite{read} model, which leads to the Klein-Gordon equation: $\left(
\partial _i+i\tilde{A}_i^h\sigma \right) ^2z^\sigma =(E/c_s)^2z^\sigma ,$
where $z^\sigma =\frac 12(\bar{b}_{A\sigma }+\bar{b}_{B\sigma }^{*}),$ $\bar{%
b}_{A\sigma }$ and $\bar{b}_{B\sigma }$ are spinons on $A$ and on $B$
sublattices, and $c_s$ is the effective spin-wave velocity\textbf{\ (}$%
c_s\sim 1.64aJ$ \cite{aa}$)$. The energy spectrum is \cite{landau,au} $%
E_{n,m}=\pm c_s\sqrt{(n+1/2+\mid m\mid \mp m)\tilde{B}^h}$ and the wave
functions are $z_{n,m}^\sigma \sim \rho ^{\mid m\mid }e^{im\varphi
}F(-n,\mid m\mid +1,\alpha ^2\rho ^2)\exp (-\alpha ^2\rho ^2/2),$ where $F$
is the hypergeometric function, $m$ is the eigen-value of the angular
momentum, and $n$ is the eigen-value of the harmonic oscillation level
number, $\alpha ^2=\tilde{B}^h/2$, $\rho =\sqrt{x^2+y^2}$. The energy gap
between first excited state and the ground state is 
\begin{equation}
E_s=E_1-E_0\sim 1.5J\sqrt{x_{eff}}.
\end{equation}
This result is in contrast to $E_s\varpropto xJ$ ($x\rightarrow 0)$ obtained
at ${<\mid }b_{i\sigma }\mid >=0$ \cite{string2}. The characteristic spin
energy is then given by $E_g\sim 2E_s$. It has been previously established 
\cite{ming,muthu} that the superconducting transition occurs when spinons
become deconfined at a finite temperature, which is determined at $T_c\simeq 
\frac{E_g}c$ with $c\sim 4$, as shown in Fig. 2.

In conclusion, the low-temperature phase diagram for a doped Mott
antiferromagnet described by (\ref{hh}) and (\ref{hs}) is basically
determined by its intrinsic topological gauge structure. A quantum critical
point at $x_c\simeq 0.043$ is found, below and above which, \emph{dual
confinement and deconfinement} take place at $T=0$, leading to a systematic
evolution from the antiferromagnetic to superconducting phases as a function
of $x$. The complex experimental phase diagram in the weakly doped cuprates
may be understood within such a framework. For example, the \emph{dipolar}
effect of doped holes is responsible for a vanishing $T_N$ at $x_0\sim 0.03$
and a cluster spin-glass phase at $x_0<x<x_c.$ The superconducting state
sets in at $x\geq x_c$, whose phase coherence is destroyed by a
deconfinement of spinons at $T\geq T_c$ \cite{ming,muthu}$.$ Due to the
space limit, we have not discussed the possible stripe instability in the
same model, which is explored at low doping as a competing phenomenon
elsewhere \cite{kou}.

\acknowledgments 
We thank T. Li and H.T. Nieh for helpful conversations. S.-P.K. is partially
supported by the NSFC Grant no. 10204004. Z.-Y.W. acknowledges partial
support from NSFC Grant no. 90103021 and no. 10247002.

\end{document}